\documentclass[conference]{IEEEtran}
\IEEEoverridecommandlockouts
\usepackage{tikz}
\usetikzlibrary{shapes,arrows,shadows,positioning,calc,math}
\usetikzlibrary{decorations.text}
\usetikzlibrary{backgrounds}
\usepackage[siunitx]{circuitikz}
\usepackage{caption}
\usepackage{subcaption}

\newtheorem{assumption}{Assumption}

\newtheorem{remark}{Remark}

\newcommand{\R}{\mathbb{R}}
\usepackage{epstopdf}
\usepackage{cite}
\usepackage{amsmath,amssymb,amsfonts}
\usepackage{algorithmic}
\usepackage{graphicx}
\usepackage{textcomp}
\usepackage{xcolor}
\def\BibTeX{{\rm B\kern-.05em{\sc i\kern-.025em b}\kern-.08em
    T\kern-.1667em\lower.7ex\hbox{E}\kern-.125emX}}
\begin{document}

\title{Separating multiscale Battery dynamics and predicting multi-step ahead voltage simultaneously through a data-driven approach\\

}

\author{\IEEEauthorblockN{Tushar Desai}
\IEEEauthorblockA{\textit{Delft Center for Systems and Control} \\
\textit{Delft University of Technology}\\
Delft, Netherlands \\
t.k.desai@tudelft.nl}
\and
\IEEEauthorblockN{Riccardo M.G. Ferrari}
\IEEEauthorblockA{\textit{Delft Center for Systems and Control} \\
\textit{Delft University of Technology}\\
Delft, Netherlands \\
r.ferrari@tudelft.nl}
}

\maketitle

\begin{abstract}
Accurate prediction of battery performance under various ageing conditions is necessary for reliable and stable battery operations. Due to complex battery degradation mechanisms, estimating the accurate ageing level and ageing-dependent battery dynamics is difficult. This work presents a health-aware battery model that is capable of separating fast dynamics from slowly varying states of degradation and state of charge (SOC). The method is based on a sequence to sequence learning-based encoder-decoder model, where the encoder infers the slowly varying states as the latent space variables in an unsupervised way, and the decoder provides health-aware multi-step ahead prediction conditioned on slowly varying states from the encoder. The proposed approach is verified on a Lithium-ion battery ageing dataset based on real driving profiles of electric vehicles.
\end{abstract}

\begin{IEEEkeywords}
battery, multiscale dynamics, machine learning
\end{IEEEkeywords}

\section{Introduction}
\label{scn:Introduction}
As the electric vehicle market and stationary energy storage systems expand, Lithium-ion (Li-ion) batteries are becoming increasingly popular due to several advantages, such as higher energy densities and extended cycle life. However, batteries experience degradation over their lifetime, which results in diminished performance. While degradation itself is due to several complex phenomena occurring at the microscale, such as solid electrolyte interphase, lithium plating and particle fracture, at the macroscopic level, this manifests as capacity fade and resistance increase. To guarantee a safe and reliable operation, the battery's health and behaviour at any particular degradation level have to be known and taken into consideration by the control system.

Model-based approaches and data-driven methods are used to estimate a battery's degradation state, which cannot be directly measured. Electrochemical (EC) models and equivalent circuit models (ECM) are two model-based approaches used for health-aware battery modeling. EC models use partial differential equations (PDEs) to capture battery degradation at a microscale. This provides greater insight but requires many parameters to be measured or estimated from data and have a high computational cost. Degradation can arise from various mechanisms, either individually or in combination. While there exist well-developed models for individual degradation mechanisms, the intricate interplay between multiple mechanisms remains poorly understood \cite{Edge2021-da}.

ECM, which is generally based on resistor-capacitor circuits, captures instead the macroscopic behaviour of the battery. ECM provides a good trade-off between accuracy and complexity and can well describe the battery behaviour at a given point in the battery lifetime and around a fixed operating condition. Still, they miss a microscopic physics foundation and cannot capture parameters' nonlinear and time-varying dependency on the State of Charge (SOC), temperature and current \cite{Lai2020-ib}. This limits the usefulness and accuracy of ECM over longer periods, making it ineffective for simulating the timescales over which degradation evolves \cite{Campestrini2017-fy}.      

Various factors affect the battery degradation phenomenon, from materials and manufacturing to working conditions  \cite{Edge2021-da}. Degradation involves various physio-chemical processes, and inferring the battery's health remains a key challenge. Due to the aforementioned limitations of physics-based models, data-driven approaches for battery health modelling are becoming increasingly attractive. Recently, there has been a significant focus on modern machine-learning (ML) methods as they can automate feature extraction from high-dimensional data and efficiently learn complex patterns from it \cite{Fink2020-ud}. These features of ML-based methods make them suitable for battery health estimation problems, and many ML approaches are proposed in the literature. 

Machine learning approaches for battery modeling can be divided into two categories. The first approach estimates battery health by using ML to estimate its capacity or resistance. For instance, Li et al.\cite{Li2021-ts} used constant current phase charging curve data to estimate capacity. Chemali et al.\cite{Chemali2018-le} used deep-learning to estimate SOC using voltage and current measurement. Aitio et al.\cite{Aitio2021-mu} inferred health and estimated End-of-life using the Gaussian process regression technique. However, most of these ML-based methods are based on supervised learning, which needs labelled data for training the models. These labels are either SOC, capacity or resistance at different degradation levels. It is challenging to get these labelled data accurately from the vehicle batteries, as they are subjected to various dynamic and varying load conditions.

The second approach of ML application in battery modeling is to develop a health-aware battery model that simulates battery behaviour over its lifetime. For instance, Zhao et al.\cite{Zhao2023-qn} and Hong et al.\cite{Hong2019-ei} used recurrent neural network (RNN) variants to do multistep-ahead voltage prediction and fault diagnosis. Firstly, the estimated SOC is used as an input signal in all these approaches, which may not always be accurate, affecting the overall model's reliability. Secondly, the limited interpretability of these models is a significant challenge due to the complex interplay of fast and slow dynamics, which makes it difficult to obtain clear insights into battery health and behaviour.

Unsupervised learning methods, which do not need labelled data for model development, could be beneficial in addressing the aforementioned issues. Among many unsupervised learning methods, autoencoder-based approaches are becoming popular for identifying nonlinear systems as reported in \cite{Gedon2021-ie, Masti2021-np, Champion2019-yh}. The ability of autoencoders to discover patterns without being explicitly trained makes them suitable for system identification. Additionally, the ease with which state-space model structures can be configured to provide interpretability further enhances their suitability for this purpose. In battery applications, autoencoders are applied for denoising signals to increase the accuracy of SOC estimation \cite{Chen2021-za} or in predicting the Remaining Useful Life (RUL) assuming labelled data of battery capacity are available \cite{Wei2022-zq}. However, directly leveraging an autoencoder's unsupervised learning ability to separate slowly varying dynamics from the fast ones and predict multi-step ahead voltage has never been attempted. 

In our work, we precisely aim to explore this research direction. By using an autoencoder's inherent capability of compressed representation learning, we build a method for inferring health and SOC in an unsupervised manner. To reach this goal, we extend the standard autoencoder approach, where input observations are reconstructed in the decoder, and instead, we build a forward prediction decoder model. The forward prediction error is used to learn the latent space representation of the battery. This data-driven battery model can infer slowly varying states in an unsupervised way and simultaneously provide accurate, ageing-aware and multi-step ahead battery voltage predictions.

Our proposed model features a 1-dimensional Convolutional Neural Network (1-D CNN)-based encoder for abstract learning of slowly varying states from the time series signals of current and voltage data. As a decoder, we employ RNN for temporal learning of ageing-aware battery dynamics, which is trained to predict the battery voltage multiple steps ahead. The decoder structure is kept shallow, with only current as external input and the latent space representation from the encoder. This not only prevents overfitting but also implicitly forces the autoencoder to learn the slow dynamics in its latent representation, as this is the only way to reduce the forward voltage prediction error.

The rest of the paper is divided into five sections as follows. General battery ageing modeling with separate timescale is detailed in Section \ref{scn:battery ageing modeling}. Section \ref{scn:model} details the model construction and the encoder-decoder-based model with details on algorithms applied. Section \ref{scn:data} details the data on which the model is validated. Lastly, the Simulation and results are discussed in Section \ref{scn:simulations}, and finally,  the Conclusion is drawn in Section \ref{scn:conclusion}.

\section{battery multiscale modeling}
\label{scn:battery ageing modeling}
The electrical dynamics of lithium-ion batteries, when either charging or discharging, have relatively fast dynamics, with time scales in the order of milliseconds to a few seconds. Conversely, degradation phenomena are relatively slow, as they manifest over several sequences of charging and discharging periods, with each pair of full charge and discharge being termed a \emph{cycle}. The time scale of degradation is thus comparable to the duration of a cycle, which is in the order of several hours  \cite{Sulzer2021-ao}. A typical battery for automotive traction has an expected lifetime of about 2000 cycles, which under typical usage corresponds to about 10-15 years of useful service life \cite{Bocca2020-df}.

The concurrent presence of fast and slow dynamics in batteries motivates us to consider a general, two-time-scale nonlinear system of the form:

\begin{equation}
	\label{eqn:twotimescale_sys}
	\begin{aligned}
	\Sigma: 
	\begin{cases}
			\dot{\xi}_{f} & = f(\xi_{f},\xi_{s}, \upsilon,\Theta),\\
                \dot{\xi}_{s} & = g(\xi_{f},\xi_{s}, \upsilon,\Theta),\\
		      \Upsilon         & = h(\xi_{f},\xi_{s},\upsilon,\Theta),
	\end{cases}
	\end{aligned}
\end{equation}

\noindent where $\xi_{f} \in \mathbb{R}^{n_{\xi f}}$ is the fast varying state vector and $\xi_{s} \in \mathbb{R}^{n_{\xi s}}$ is the slowly varying one. $\upsilon \in \mathbb{R}^{n_{\upsilon}}$ is a controlled input vector and $\Upsilon \in \mathbb{R}^{n_{\Upsilon}}$ is a measured output vector. Finally, $\Theta \in \mathbb{R}^{n_{\Theta}}$ is a parameter vector which accounts for the battery specific technology and for unit-to-unit variations.

Systems with distinct timescales can be modeled and controlled using, for instance, the asymptotic method \cite{Kevorkian1996-qm, Khalil2002-zn}. When exact and precise equations for a particular dynamic are not available, the problem at hand becomes increasingly challenging. In many complex systems, multiple timescales are intertwined, making it difficult to accurately separate them. This issue is also observed in battery models, where the exact timescale separation remains unknown, leading to a scarcity of attempts in the literature to consider models with an explicit multi-timescale separation.

In the following sections, we show how this impasse can be overcome using a data-driven approach based on training an autoencoder. Autoencoders have been successfully applied to approximate physical phenomena from high dimensional data and accurately predict future outputs \cite{Wiewel2019-ap}. In the present paper, we use an autoencoder to separate slowly varying SOC and ageing states from fast ones and simultaneously predict the battery output over multiple time steps ahead. The structure of our proposed autoencoder is explained next.
\section{Model construction}
\label{scn:model}
The standard autoencoder model is designed for signal reconstruction, where the encoder compresses high dimensional input observations to a reduced \emph{latent space}, which is then used by the decoder to reconstruct the given input observations. This structure has been widely applied for instance for anomaly detection, where the deviation of the reconstruction error from the normal behaviour is used as a measure to distinguish anomaly. As mentioned by \cite{Masti2021-np}, this standard approach generates a so-called stiff latent space, which is not conducive to controller or observer design. To avoid this the autoencoder can instead be designed for predicting one or multiple steps ahead the output signal, instead of merely reconstructing its actual value.     

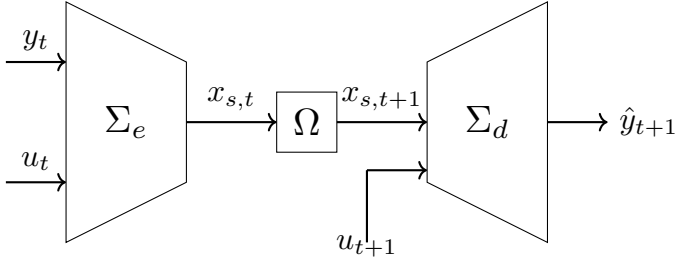
\begin{figure}[h!]
\begin{center}
    \begin{circuitikz}[scale = 0.8, transform shape]
       \draw (0,0) -- (0,4) -- (2,3) -- (2,1) -- cycle ;
       \draw [white, very thick, |-] (0.5,2) -- (1.5,2) node[midway ,align=center, scale=1.8, black] {$\Sigma_e$};
       
       \draw (3.5,1.5) -- (3.5,2.5) -- (4.5,2.5) -- (4.5,1.5) -- cycle;
        \draw [white, very thick, |-] (3.6,2) -- (4.4,2) node[midway ,align=center, scale=1.8, black] {$\Omega$};
       
       \draw (6,1) -- (6,3) -- (8,4) -- (8,0) -- cycle;
        \draw [white, very thick, |-] (6.5,2) -- (7.5,2) node[midway ,align=center, scale=1.8, black] {$\Sigma_d$};

       \draw[thick,->] (-1,3) -- (0,3) node[midway, above, scale=1.5] {$y_{t}$};
       \draw[thick,->] (-1,1) -- (0,1) node[midway, above, scale=1.5] {$u_{t}$};
       \draw[thick,->] (2,2) -- (3.5,2) node[midway, above, scale=1.5] {$x_{s,t}$};
       \draw[thick,->] (4.5,2) -- (6,2)  node[midway, above, scale=1.5] {$x_{s,t+1}$};
       \draw(5,0) -- (5,1.2)node[near start, below, scale=1.5] {$u_{t+1}$}[thick,->] (5,1.2) -- (6,1.2);
       \draw[thick,->] (8,2) -- (9,2) node[right, scale=1.5] {$\hat{y}_{t+1}$};
    \end{circuitikz}
\end{center}
\caption{\centering Battery Autoencoder model }
\label{fig:autoencoder}
\end{figure}

Based on the physical insight of the problem at hand, we propose the autoencoder structure depicted in Figure~\ref{fig:autoencoder}, where $\Sigma_e$ represents the encoder and $\Sigma_d$, the decoder. The symbol $y \in \mathbb{R}^{n_y}$ denotes the battery output, that is, its voltage, while $u \in \mathbb{R}^{n_u}$ is an input vector consisting of the battery current. We consider a discrete time domain, with the index $t$ denoting the $t^{th}$ sampling period. The encoder output, denoted by $x_{s} \in \mathbb{R}^{n_{xs}}$, is the reduced latent variable which, as a result of our design choice for the autoencoder structure, encodes the present battery slowly varying states.

The function $\Omega: \mathbb{R}^{n_{xs}} \mapsto \mathbb{R}^{n_{xs}}$ is the state transition function for $x_{s}$, which thus represents the evolution over a single time step of the battery slow states. Finally, the symbol $\hat{y} \in \mathbb{R}^{n_y}$ denotes the one time step ahead prediction of the battery output computed by the decoder.

       
       


Formally, the autoencoder model can be defined through the following set of discrete time equations:

\begin{subequations}
    \label{eqn:autoen_model}
    \begin{align}
        &\Sigma_e:
        \begin{cases}
            \begin{aligned}
                x_{s, t}    &= g_{w}(u_{t},y_{t}),  \\
	            x_{s, t+1}  &= \Omega(x_{s, t}),
            \end{aligned}
        \end{cases} \\
        &\Sigma_d:     
        \begin{cases}
            \begin{aligned}
                x_{f, t+1}     &= f_{w}(x_{f, t},\:  x_{s, t},\:  u_{t}),  \\
	            {\hat{y}}_{t+1} &= h_{w}(x_{f, t+1},\: x_{s, t+1},\: u_{t+1}),
            \end{aligned}
        \end{cases}
    \end{align}
\end{subequations}



\noindent where $w \in \mathbb{R}^{n_w}$ is the tuneable weights vector associated with $\Sigma_e$ and $\Sigma_d$. The vector $x_{f} \in \mathbb{R}^{n_{xf}}$ is the fast varying state, while the functions $f_w$, $g_w$ and $h_w$ are parameterized in $w$ and are meant to approximate the fast dynamics $f$, the slow dynamics $g$ and the output function $h$ which were defined in ~\eqref{eqn:twotimescale_sys}. These choices are introduced to show explicitly how, in the proposed structure, the encoder aims to learn the slow $x_{s}$ while the decoder aims to learn the fast $x_{f}$ and the dependency of $x_{f}$ on $x_{s}$.



\begin{assumption}
    In the present paper we assume that the discrete sampling time is comparable to the fast dynamics time scale, such that the approximation $x_{s,t+1} = \Omega(x_{s,t}) \approx x_{s,t}$ can be done.
\end{assumption}

\begin{remark}
     Eq.~\eqref{eqn:autoen_model} is based on using one single sample of $y_t$ and $u_t$ in the past to predict one sample $\hat y_{t+1}$ in the future based on the knowledge of $u_{t+1}$. The scheme can be extended to the case where $n_a$ samples in the past are used to predict $n_b$ samples in the future, in order to implement the multiple step ahead prediction mentioned earlier. For the sake of ease of notation and brevity, in the following we will not explicitly denote the case with $n_a>1$ and $n_b>1$, unless needed.
\end{remark}

To effectively capture the slowly varying state, the encoder mapping $g_w$ should maximise the autocorrelation of $x_s$ between two consecutive timesteps in any given cycle $k$. Thus, the training of the autoencoder can be obtained by solving the following optimization problem numerically:

\begin{equation} \label{eq:cost_function}
    \begin{aligned}
    \arg\min _{w}  \mathcal{L}_{pred}(w)  + \lambda \mathcal{L}_{corr}(w)\\
   \text{s.t.} \textrm{\eqref{eqn:autoen_model}} \\
\end{aligned}
\end{equation}

\noindent where  $\mathcal{L}_{pred}(w) =  \sum_{k=0}^{K}\sum_{t=0}^{T} L(y^k_{t+1},\hat{y}^k_{t+1})$ and $L: \R^{n_y} \times \R^{n_y} \mapsto \R^+$ is any suitable loss function, such as for instance the mean square error. The superscript $k$ is used to identify a given charging and discharging cycle, with $k=0$ denoting a fresh battery and $k$ increasing monotonically during the battery lifetime. Furthermore, each cycle k lasts for a certain time period T. $\mathcal{L}_{corr}(w)$ is the autocorrelation loss function between $x_{s,t-1}$ and $x_{s,t}$. $\lambda$ in \eqref{eq:cost_function} is the correlation regularization hyperparameter.

\begin{equation} \label{eq:correlation_cost_function}
    \begin{aligned}
        \mathcal{L}_{\text {corr}}=- \left(\sum_{i=j}^{n_{xs}}\left|r_{i, j}\right| \right)
    \end{aligned}
\end{equation}
where,

\begin{equation} \label{eq:correlation_function}
    \begin{aligned}
        r_{i, j}=\frac{1}{K} \sum_{k=0}^K \frac{\sum_t^T\left(x^{k,i}_{s,t}-\bar{x}^{k,i}_{s}\right)\left(x^{k,j}_{s,t-1}-\bar{x}^{k,j}_{s}\right)}{\sqrt{\sum_t^T\left(x^{k,i}_{s,t}-\bar{x}^{k,i}_{s}\right)^2 \sum_t^T\left(x^{k,j}_{s,t-1}-\bar{x}^{k,j}_{s}\right)^2}}
    \end{aligned}
\end{equation}

\noindent i.e, $r_{i,j}$ represents the correlation between the latent space values of features $i$ and $j$ at two consecutive timesteps $t$ and $t-1$. $\bar{x}^{k,i}_{s}$ is the mean value of the $i^{th}$ latent space feature over cycle $k$. 





We now outline the deep learning algorithms employed in the model's encoder and decoder parts for implementing the functions $f_w$, $g_w$ and $h_w$.

\subsection{Encoder}
\label{subscn:1D-convolution}

Our proposed encoder is based on a Convolutional Neural Network (CNN). CNNs are capable of learning complex patterns and objects from 2-D signal data, e.g. videos and images \cite{Gu2018-dq}. CNN variants for 1-D sensor data, known as 1-D CNN, have seen increasing applications in the domain of signal processing for anomaly detection and structural health monitoring \cite{Kiranyaz2021-hy}. Compared to Gated Recurrent Units (GRU), described in \ref{subscn:RNN}, 1-D CNN can effectively handle longer sequences of data with less computation burden and also offers better abstract learning \cite{Fink2020-ud}. These features make it suitable for the slowly-varying state extraction application of the encoder, as it is essential to have a longer receptive field to observe significant variations.

\begin{figure}[h!]
    \centering
    \includegraphics[width=1\linewidth]{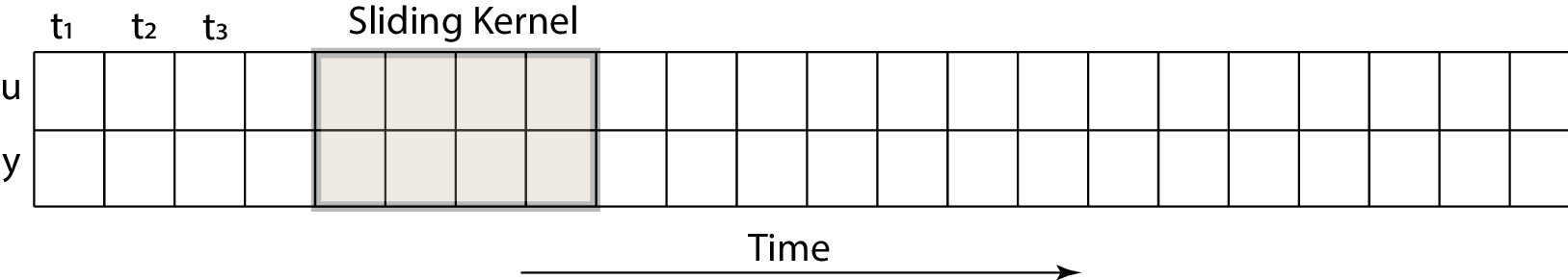}
    \caption{\centering 1-D CNN with sliding kernel on encoder input data}
    \label{fig:1-D_CNN}
\end{figure}

In 1D convolution, discrete linear operations are performed on encoder input data, denoted here generically as $H_t \in \mathbb{R}^{n_u + n_y}$, $t \in \{1,\ldots 1+a\}$, with $m$ sliding kernels of size $L_w$ for input channels $n=n_u + n_y$. $W_{ij} \in \mathbb{R}_{L_w}$, $i \in \{1,\ldots, n\}$ and $j \in \{1,\ldots,m\}$  are the weights associated with kernels. This convolution operation results in the output vector $q_j \in \mathbb{R}^{n_{out}}$, $j \in \{1,\ldots,m\}$ with $n_{out} = a - L_w +1$. Formally, the 1-D convolution operation can be described as:

\begin{equation}
\begin{aligned}
    q_j &=\sum_{i=1}^n \operatorname{Conv1D}\left(H_i, W_{(i, j)}\right) \\
    q_j[k] &=\sum_{i=1}^n \sum_{r=1}^{L_W} X_i[k+r-1] W_{(i, j)}[r]
\end{aligned}
\end{equation}

By cascading multiple convolution operations with various kernel sizes $L_w$, the dimension $n_{out}$ of the CNN output can be made to reach a given latent space dimension $n_{xs}$, which is a hyperparameter. This multiple 1D-CNN layers-based deep encoder model can then be used to estimate $x_{s,t}$ and provide it as an input to the decoder.

\subsection{Decoder}
\label{subscn:RNN}
The decoder we propose is based on Recurrent Neural Networks (RNN). RNNs are the generalisation of a feedforward artificial neural network (ANN) to data sequences. Among the RNN variants, the Gated Recurrent Unit (GRU) was proposed by  \cite{Chung2014-yo} to force the recurrent unit to capture adaptive dependencies of different timescales with fewer computations compared to  Long-Short Term Memory (LSTM, \cite{Hochreiter1997-uf}).

In particular, we use the GRU-based single-layer decoder model, followed by a time-distributed, single neuron feedforward ANN layer for learning the $x_{s}$ dependent battery fast dynamics $f_w$ and the output $h_w$. A GRU-based decoder model can be described in state-space form as:

\begin{equation}
\label{eqn:GRU}
\begin{aligned}
    \begin{cases}
        X_{t+2} &= z_{t+1}  \tanh \left(W_r u_{t+1}+U_r (r_{t+1} \circledcirc X_{t+1})+b_r\right)+ \\
             &\hspace{.4cm}( 1-z_{t+1})  X_{t+1} \\
        z_{t+1}  &=\sigma\left(W_z u_{t+1} + U_z X_{t+1}+b_z\right) \\
        r_{t+1} &=\sigma\left(W_f u_{t+1} +U_f X_{t+1}+b_f\right) \\
        \hat{y}_{t+1} &= \tanh (W_0 X_{t+1}+b_0).
    \end{cases}    
\end{aligned}
\end{equation}

\begin{figure}[h!]
    \centering
    \includegraphics[width=0.8\linewidth]{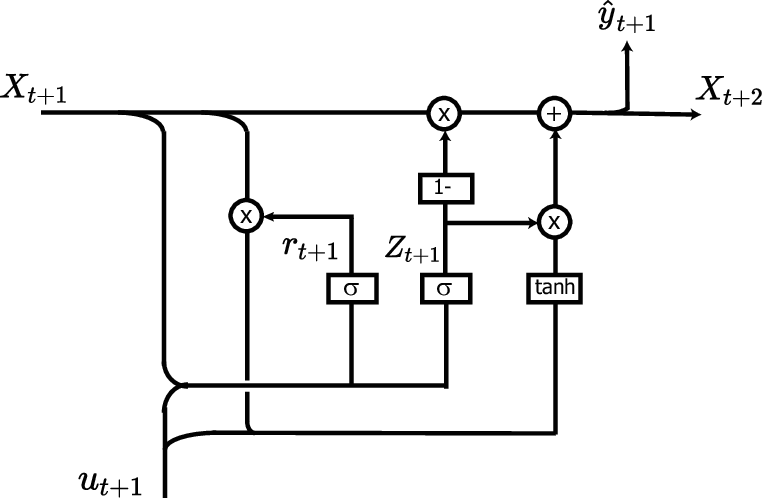}
    \caption{\centering GRU unit with reset $(r_{t+1})$ and update $(z_{t+1})$ gates}
    \label{fig:GRU}
\end{figure}

The GRU cell state vector $X \in \mathbb{R}^{nx}$  carries the fast and slowly varying state info across the decoder's future timesteps. For the first timestep, $X_{t+1} = x_{s, t+1}$, as the decoder acquires $x_{s}$ from the encoder. For future timesteps, $X_{t+i} \: (i > 1)$ carries a fast varying state depending on the inputs and also $x_{s}$ acquired from the encoder. $W, U$ and $b$ are the weights and bias parameters associated with the update gate ($z_t$) and reset gate ($r_t$). Furthermore, the symbol $\circledcirc$ indicates element-wise multiplication, while. $\sigma$ (sigmoid) and $\tanh$ are nonlinear activation functions.
\section{Data}
\label{scn:data}
\begin{figure}[h!]
    \centering
    \includegraphics[trim={1.1cm 0.17cm 0.3cm 0},clip,scale=0.41]{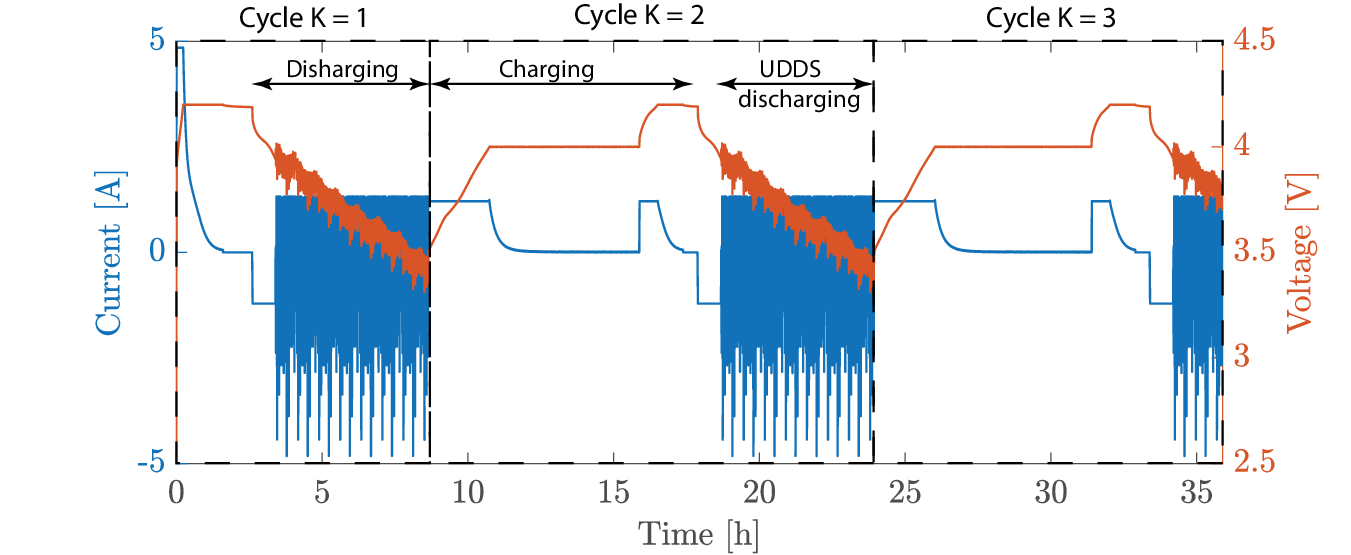}
    \caption{\centering Battery cycling data with CCCV charging and CC-UDDS discharge profile \cite{Pozzato2022-ad}}
    \label{fig:stanford_udds}
\end{figure}

To validate the proposed autoencoder approach, an open-source Lithium-ion battery ageing dataset provided by \cite{Pozzato2022-ad} is used. The dataset consists of battery cells with a nominal capacity of $\SI{4.85}{Ah}$ ($Q_{nom}$), which were cycled for 23 months using a Constant Current (CC)-Constant Voltage (CV) charging protocol with varying charging rates ($C/4$ to $3C$), and the real Urban Dynamometer Driving Schedule (UDDS) based discharge driving profile, as shown in Figure~\ref{fig:stanford_udds}. To provide a ground truth for the health of the batteries, various Characterisation tests, including capacity, Hybrid Pulse Power Characterization (HPPC), and Electrochemical Impedance Spectroscopy (EIS), were performed intermittently at intervals of $25-30$ cycles.

\begin{figure}[h!]
    \centering
    \includegraphics[trim={0cm 0cm 0cm 1.48cm},clip,width=1\columnwidth]{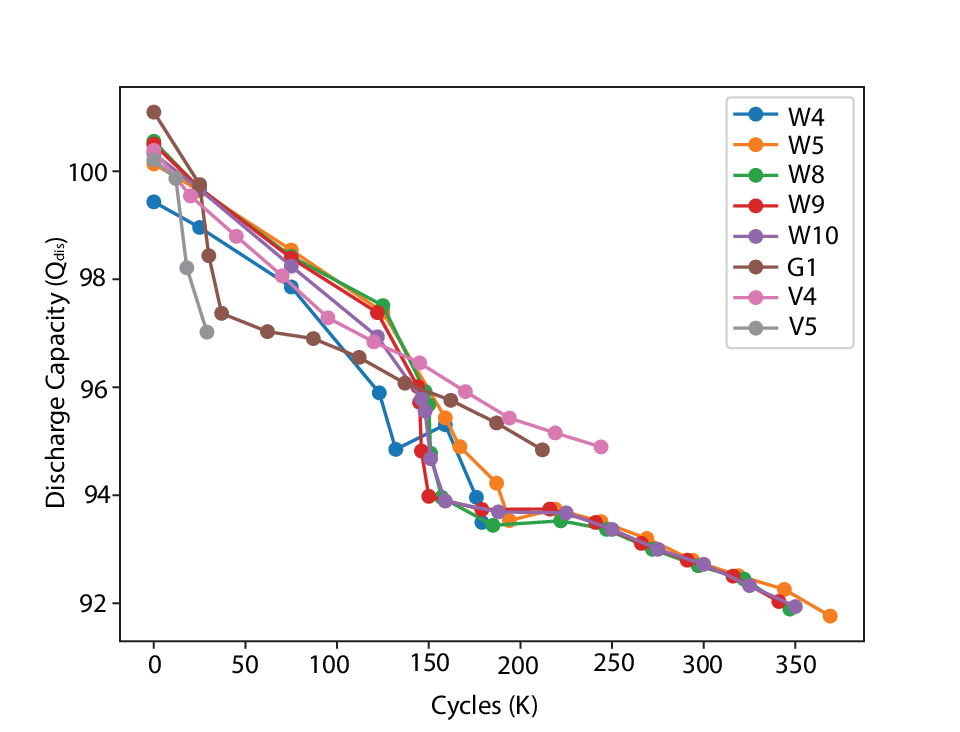}
    \caption{\centering Varying Battery discharge capacity measured intermittently across different cycles  \cite{Pozzato2022-ad}}
    \label{fig:discharge_capacity}
\end{figure}

The cell's discharged capacity is evaluated through a capacity test conducted by discharging it at a $C/20$ rate from a fully charged state. The discharge capacity of the cells over the cycles is reduced due to ageing, as shown in Figure~\ref{fig:discharge_capacity}, and is calculated using:

\begin{equation}
    \label{eq:discharge_capacity}
    Q_{dis} = {\frac{\int I(t) \,dt}{Q_{nom} \times 3600} \times 100 ,}
\end{equation}

\noindent where the discharge current $I(t)= 0.24A$ used during the capacity test is integrated over the test period.

We used eight battery cells' UDDS discharge profile data for our model development (cell labels shown as a legend in Figure~\ref{fig:discharge_capacity}). The battery is discharged from $\SI{80}{\%}$ SOC to $\SI{20}{\%}$ SOC with multiple UDDS cycles. The data from a cell $G1$ are kept aside for model testing, while the other seven cell data are split into $\SI{80}{\%}$ training set and $\SI{20}{\%}$ validation set during the training phase to prevent model overfitting. All data are acquired at a frequency of \SI{10}{Hz}. 

Figure~\ref{fig:discharge_capacity} shows the varying battery discharge capacity measured intermittently across different cycles, providing a clear understanding of the battery ageing behaviour. As the temperature is kept constant during each cycle, only current and voltage signals are used for model development. To accelerate the training process, all data are normalized.

\section{Simulations and results}
\label{scn:simulations}

 After initial model training with a GRU-based encoder-decoder on a given dataset, it is inferred that latent space dimension $n_{xs}$, encoder's input history length $n_a$ and decoder's prediction horizon $n_b$ are the three primary hyperparameters to tune. These hyperparameters affect the encoder's capability to capture $x_{s}$ and ultimately impact the decoder's prediction performance. The software developed is based on the TensorFlow deep learning library. Computation operations are carried out on a cluster computer equipped with an AMD EPYC 7402 (24 Cores, 2.80 GHz) processor, 200 GB of RAM, and an Nvidia Tesla V100S-PCIE-32GB GPU. 

\begin{figure}[h!]
    \centering
    \includegraphics[trim={0cm 0 28cm 0},clip,width=0.85\linewidth]{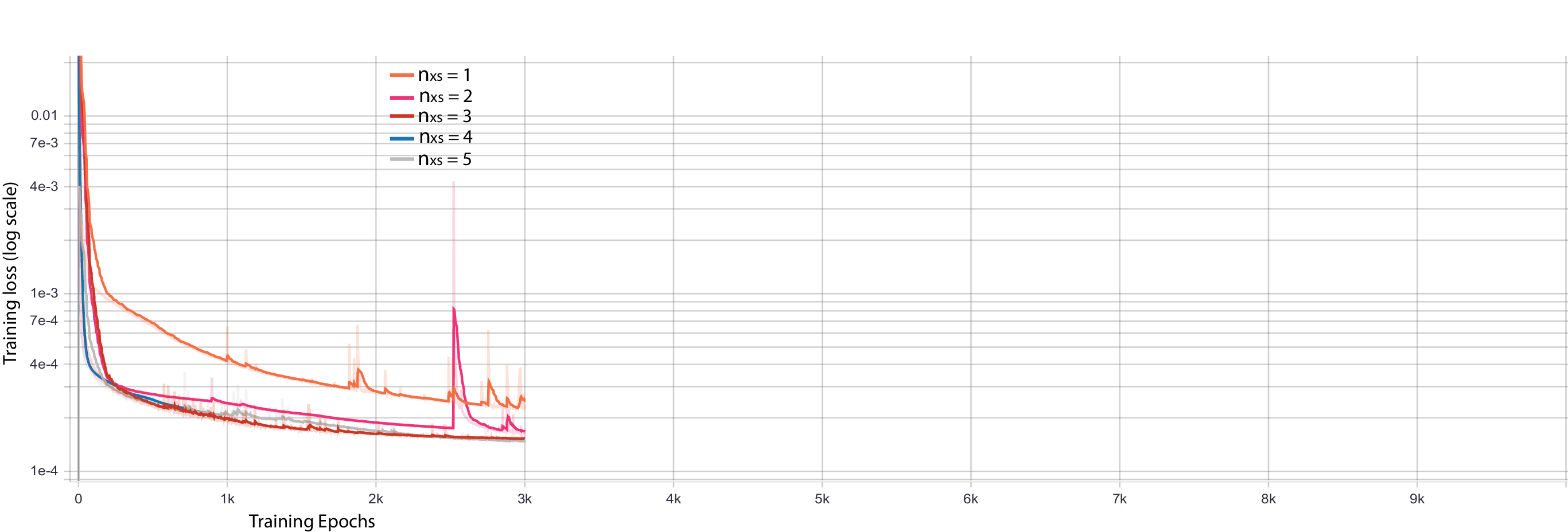}
    \caption{\centering Model training MSE loss vs epochs for $n_{xs} = 1$ to $n_{xs} = 5$ for encoder-decoder model}
    \label{fig:Training_loss_with_nxs}
\end{figure}

It is essential to optimally select the state dimension $n_{xs}$ in order to reduce the multi-step ahead prediction error. Too low an $n_{xs}$ hinders the encoder training, as weights are updated based on the chain rule of derivative across the timesteps of the encoder and decoder. On the contrary, too high an $n_{xs}$ is redundant and does not improve the decoder's prediction performance either, as seen from Figure~\ref{fig:Training_loss_with_nxs}. A campaign of simulation experiments was performed to select the suitable hyperparameter $n_{xs}$. The Mean Squared Error (MSE) loss as a function of training epochs is shown in Figure~\ref{fig:Training_loss_with_nxs} for different values of $n_{xs}$: from the Figure~\ref{fig:Training_loss_with_nxs} we can show that the value $n_{xs} = 1$ restricts the model training, while increasing $n_{xs} > 3$ does not improve the MSE significantly, and hence $n_{xs} = 3$ is ultimately determined.

GRU learning is based on the backpropagation through time algorithm  \cite{Werbos1990-uu}, which becomes computationally expensive with longer sequences due to its sequential nature of gradient updates.
On the other side, learning on shorter sequences hinders the encoder's ability to capture $x_{s}$ accurately. To find a tradeoff between robust health learning and computational burden, we initially set $n_a$ and $n_b$ to $200$. Although the decoder was able to predict the voltage accurately, the latent space output from the encoder was inconsistent. It was reasoned that the encoder may not have had enough context to accurately infer the health with short-horizon history. Increasing the $n_a$ alleviates inconsistency in the latent space but results in a higher computational burden and a vanishing gradient problem. 1D-CNN, described in \ref{subscn:1D-convolution}, is better at abstract learning and more efficient at handling longer sequences without any significant computational burden than GRU. Selecting $n_a=500$ with a 1-D CNN-based encoder provided consistent health estimation.

Finally, after setting the encoder-decoder architectures based on 1D-CNN and GRU, respectively, and setting $n_{xs}=3$, $n_a=500$, and $n_b=200$, the model was trained on the dataset mentioned in Section \ref{scn:data}. Model test results on the $G1$ battery test dataset are described here.

\subsection{Latent space}
\label{subscn:latent_space}

\begin{figure}[h!]
    \centering
    \includegraphics[width=1\linewidth]{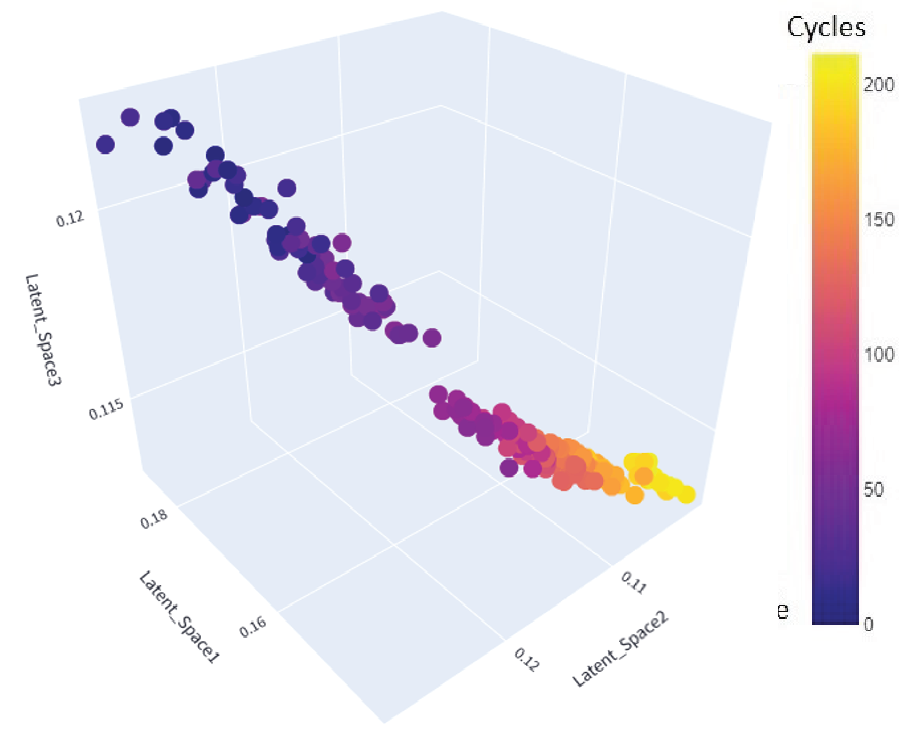}
    \caption{\centering Battery Encoder model's latent space evolution across cycles for constant SOC=$80\%$}
    \label{fig:Cell_latent_space}
\end{figure}

Figure~\ref{fig:Cell_latent_space} illustrates the reduced latent space output from the encoder across various cycles while holding the state of charge (SOC) constant. Notably, a distinct pattern emerges in the latent space as the fresh battery ages with charging and discharging cycles. To verify the encoder model's effectiveness in capturing the ageing state, data from a previously unseen cell replaces the original holdout dataset. The consistent emergence of patterns with ageing across different cells establishes the encoder's ability to track the gradual shift in battery health level when provided with a sufficient history of input-output data. The gradual shift in the latent space is similar to the gradually changing discharge capacity ($Q_{dis}$) shown in Figure~\ref{fig:discharge_capacity}. This result highlights the potential of the encoder model in accurately capturing the gradual degradation of battery health over time.

\begin{figure}[h!]
    \centering
    \includegraphics[width=1\linewidth]{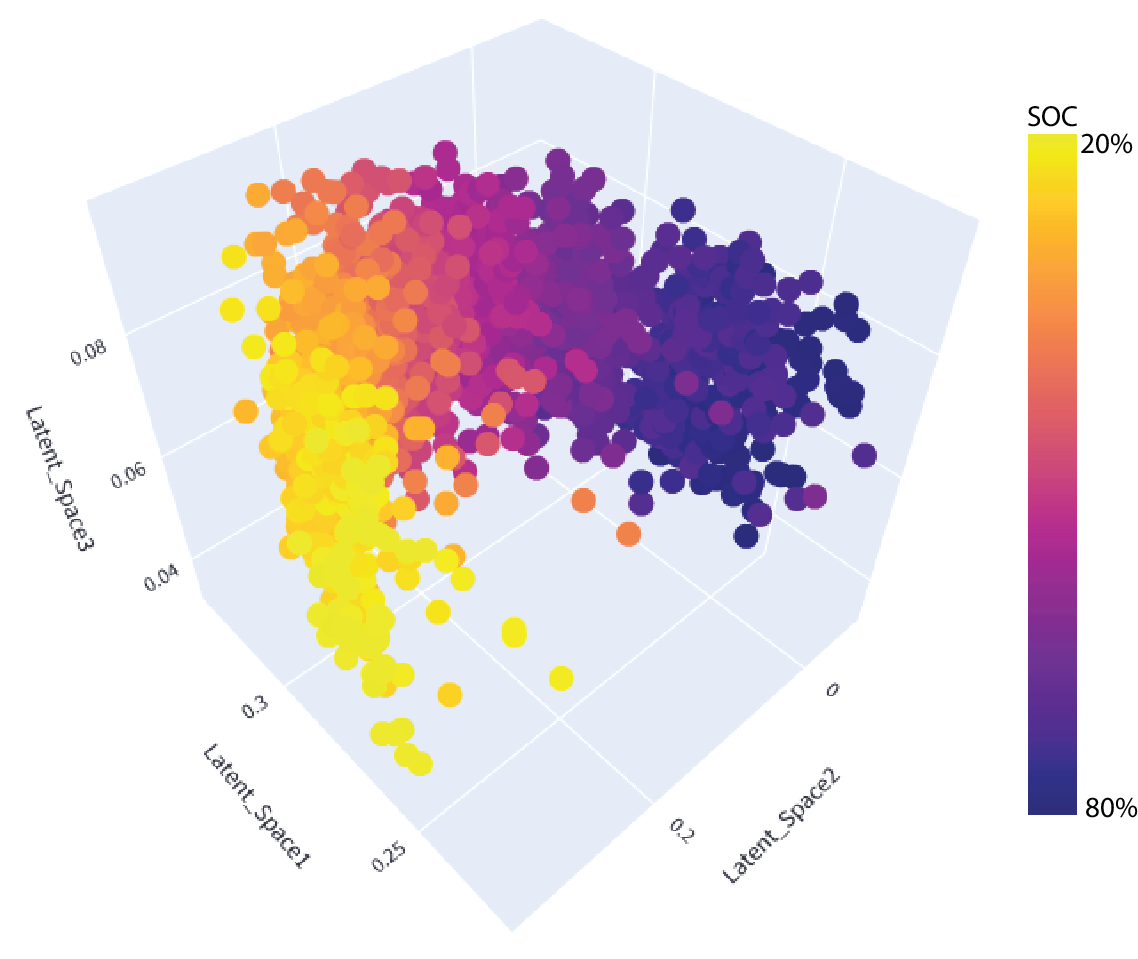}
    \caption{\centering Evolving latent space within one single discharge cycle as SOC changes from $80\%$ to $20 \%$}
    \label{fig:SOC_in_latent}
\end{figure}

Similarly, the encoder model can also effectively capture the SOC within a single cycle of battery operation. This can be observed in the latent space output for different SOC levels within a single discharge cycle, as shown in Figure~\ref{fig:SOC_in_latent}. The separation between the SOC levels in the latent space confirms the ability of the encoder to capture the SOC. This proves the encoder model's capability of capturing slowly varying SOC and ageing state without being explicitly trained for and also proves the autoencoder's ability to separate $x_s$ from $x_f$.

\subsection{Health-aware voltage prediction}
\label{subscn:prediction}
\begin{figure}[h!]
    \centering
        \begin{subfigure}[b]{0.55\textwidth}
        \includegraphics[width=0.90\linewidth]{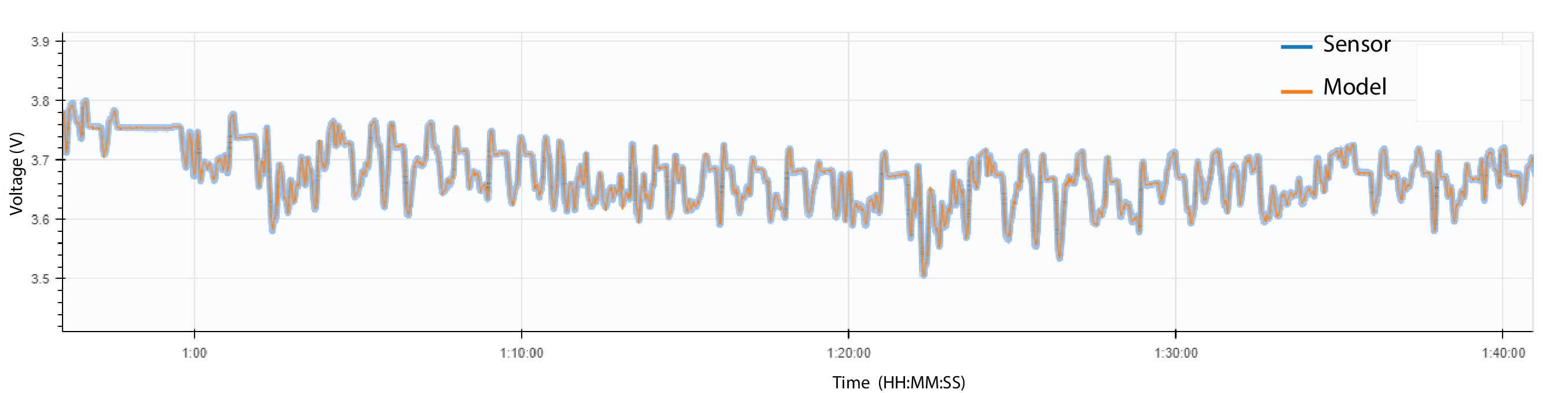}
        \caption{}
        \label{fig:multistep_ahead_sensor_model}
    \end{subfigure}
    \begin{subfigure}[b]{0.55\textwidth}
        \includegraphics[width=0.90\linewidth]{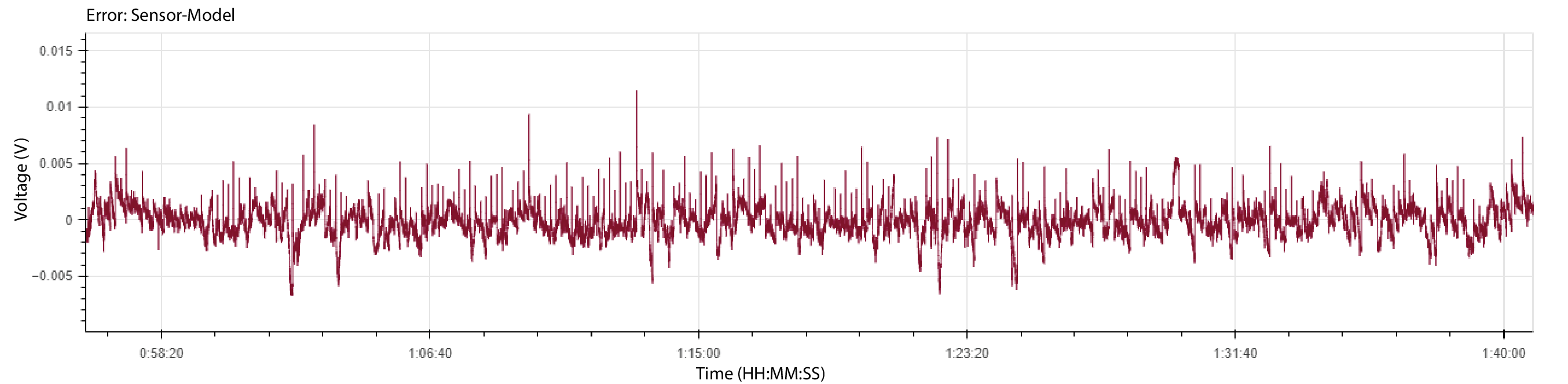}
        \label{fig:multistep_ahead_error}
        \caption{}
    \end{subfigure}
\caption{\centering (a) Multistep ahead voltage prediction and sensor measurement on test cell's $212^{th}$ UDDS cycle data (b) Error (Sensor-Model) in prediction}
\label{fig:multistep_ahead_prediction}
    
\end{figure}

\begin{figure}[h!]
    \centering
    \includegraphics[width=1\columnwidth]{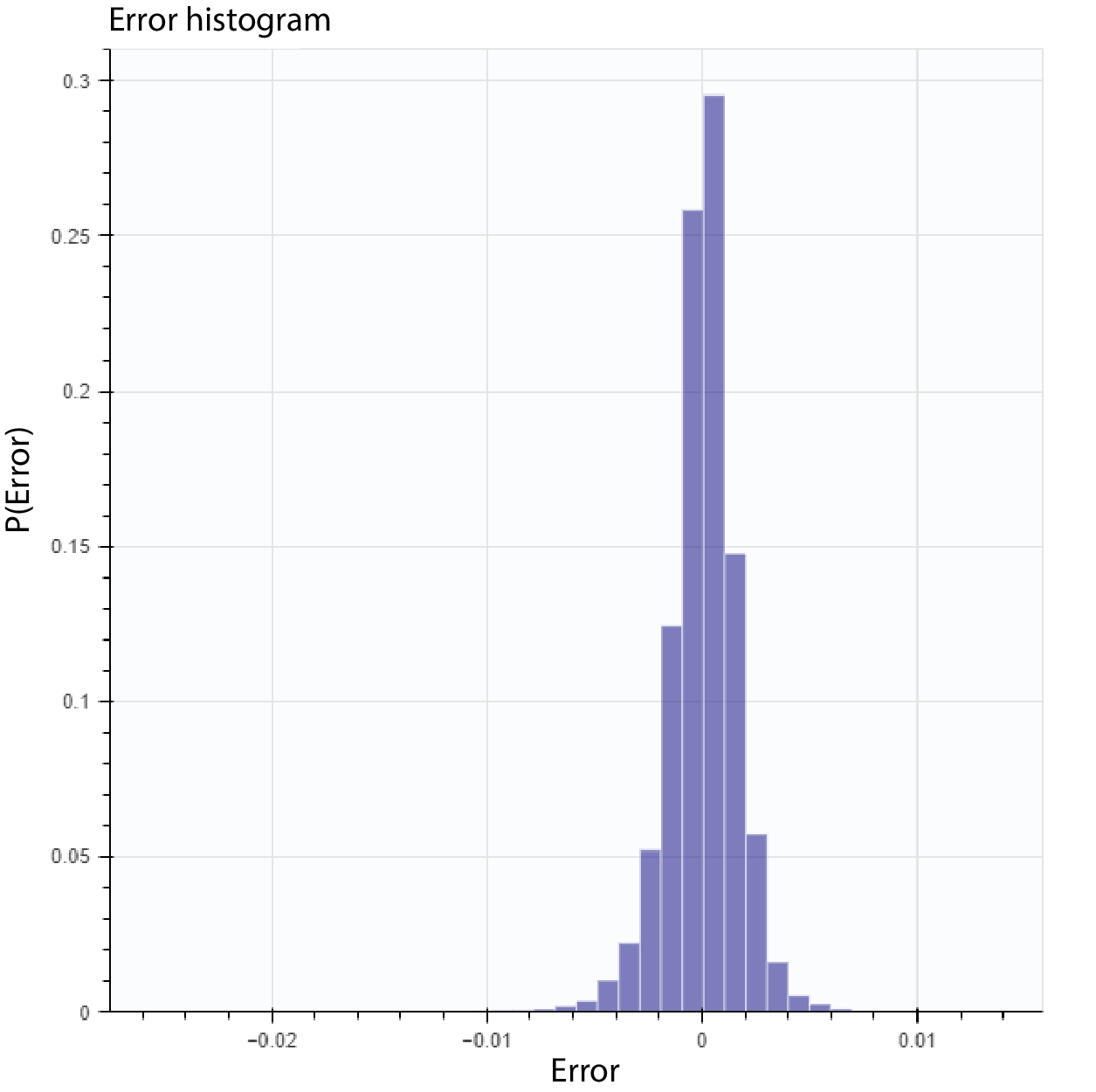}
    \caption{\centering Error histogram of the Multistep ahead voltage prediction on test cell's all ($212$) UDDS cycle data }
    \label{fig:multistep_ahead_hist}
\end{figure}

The results presented in Figure~\ref{fig:multistep_ahead_prediction} and Figure~\ref{fig:multistep_ahead_hist} demonstrate the effectiveness of the proposed decoder model in accurately predicting the voltage of a battery over a range of state-of-charge values and ageing conditions. Specifically, the decoder model predicted the voltage for a total of $n_b=200$ timesteps, corresponding to a prediction horizon of 20 seconds, with a sampling frequency of $\SI{10}{\hertz}$. To evaluate the accuracy of the predictions, two metrics were used - the root mean squared error (RMSE) and maximum absolute error (MAE). The RMSE and MAE values were found to be $\SI{51}{\milli\volt}$ and $\SI{60}{\milli\volt}$, respectively, across all 212 discharge cycles of the test cell. These results suggest that the decoder model is effective in capturing the ageing and SOC-dependent dynamics of the battery, demonstrating its ability to accurately predict the battery's performance under varying operating conditions.

\section{Conclusion}
\label{scn:conclusion}

We presented an unsupervised learning approach to separate the multiscale dynamics of batteries, which involves separating the slowly varying ageing and state of charge (SOC) states from the fast cycling dynamics. Our approach involves reconfiguring the standard autoencoder structure. We evaluated the proposed approach using a real driving profile dataset, and the results demonstrate that the model's encoder accurately captures the ageing and SOC states, while the decoder provides accurate multi-step ahead predictions. 
In contrast to other machine learning-based battery models, the time-scale separation-based approach presented here offers enhanced interpretability by isolating ageing dynamics from fast dynamics. This strategy provides a foundation for developing more precise, physics-informed machine learning models. As the fast dynamics decoder component can incorporate a physics-based model structure, which further improves the accuracy and interpretability of the model. Future work involves mapping latent space into an accurate estimation of the state of charge and health state and incorporating uncertainty in the multi-step ahead voltage predictions to quantify the confidence level.   

\bibliographystyle{IEEEtran}
\bibliography{paperpile}

\begin{thebibliography}{10}
\providecommand{\url}[1]{#1}
\csname url@samestyle\endcsname
\providecommand{\newblock}{\relax}
\providecommand{\bibinfo}[2]{#2}
\providecommand{\BIBentrySTDinterwordspacing}{\spaceskip=0pt\relax}
\providecommand{\BIBentryALTinterwordstretchfactor}{4}
\providecommand{\BIBentryALTinterwordspacing}{\spaceskip=\fontdimen2\font plus
\BIBentryALTinterwordstretchfactor\fontdimen3\font minus \fontdimen4\font\relax}
\providecommand{\BIBforeignlanguage}[2]{{%
\expandafter\ifx\csname l@#1\endcsname\relax
\typeout{** WARNING: IEEEtran.bst: No hyphenation pattern has been}%
\typeout{** loaded for the language `#1'. Using the pattern for}%
\typeout{** the default language instead.}%
\else
\language=\csname l@#1\endcsname
\fi
#2}}
\providecommand{\BIBdecl}{\relax}
\BIBdecl

\bibitem{Edge2021-da}
J.~S. Edge, S.~O'Kane, R.~Prosser, N.~D. Kirkaldy, A.~N. Patel, A.~Hales, A.~Ghosh, W.~Ai, J.~Chen, J.~Yang, S.~Li, M.-C. Pang, L.~Bravo~Diaz, A.~Tomaszewska, M.~W. Marzook, K.~N. Radhakrishnan, H.~Wang, Y.~Patel, B.~Wu, and G.~J. Offer, ``\BIBforeignlanguage{en}{Lithium ion battery degradation: what you need to know},'' \emph{\BIBforeignlanguage{en}{Phys. Chem. Chem. Phys.}}, vol.~23, no.~14, pp. 8200--8221, Apr. 2021.

\bibitem{Lai2020-ib}
X.~Lai, S.~Wang, S.~Ma, J.~Xie, and Y.~Zheng, ``Parameter sensitivity analysis and simplification of equivalent circuit model for the state of charge of lithium-ion batteries,'' \emph{Electrochim. Acta}, vol. 330, p. 135239, Jan. 2020.

\bibitem{Campestrini2017-fy}
C.~Campestrini, S.~Kosch, and A.~Jossen, ``Influence of change in open circuit voltage on the state of charge estimation with an extended kalman filter,'' \emph{Journal of Energy Storage}, vol.~12, pp. 149--156, Aug. 2017.

\bibitem{Fink2020-ud}
O.~Fink, Q.~Wang, M.~Svens{\'e}n, P.~Dersin, W.-J. Lee, and M.~Ducoffe, ``Potential, challenges and future directions for deep learning in prognostics and health management applications,'' \emph{Eng. Appl. Artif. Intell.}, vol.~92, p. 103678, Jun. 2020.

\bibitem{Li2021-ts}
W.~Li, N.~Sengupta, P.~Dechent, D.~Howey, A.~Annaswamy, and D.~U. Sauer, ``Online capacity estimation of lithium-ion batteries with deep long short-term memory networks,'' \emph{J. Power Sources}, vol. 482, p. 228863, Jan. 2021.

\bibitem{Chemali2018-le}
E.~Chemali, P.~J. Kollmeyer, M.~Preindl, and A.~Emadi, ``State-of-charge estimation of li-ion batteries using deep neural networks: A machine learning approach,'' \emph{J. Power Sources}, vol. 400, pp. 242--255, Oct. 2018.

\bibitem{Aitio2021-mu}
A.~Aitio and D.~A. Howey, ``Predicting battery end of life from solar off-grid system field data using machine learning,'' \emph{Joule}, 2021.

\bibitem{Zhao2023-qn}
H.~Zhao, Z.~Chen, X.~Shu, J.~Shen, Y.~Liu, and Y.~Zhang, ``Multi-step ahead voltage prediction and voltage fault diagnosis based on gated recurrent unit neural network and incremental training,'' \emph{Energy}, vol. 266, p. 126496, Mar. 2023.

\bibitem{Hong2019-ei}
J.~Hong, Z.~Wang, and Y.~Yao, ``Fault prognosis of battery system based on accurate voltage abnormity prognosis using long short-term memory neural networks,'' \emph{Appl. Energy}, vol. 251, p. 113381, Oct. 2019.

\bibitem{Gedon2021-ie}
D.~Gedon, N.~Wahlstr{\"o}m, T.~B. Sch{\"o}n, and L.~Ljung, ``Deep state space models for nonlinear system identification,'' \emph{IFAC-PapersOnLine}, vol.~54, no.~7, pp. 481--486, Jan. 2021.

\bibitem{Masti2021-np}
D.~Masti and A.~Bemporad, ``Learning nonlinear state--space models using autoencoders,'' \emph{Automatica}, vol. 129, p. 109666, Jul. 2021.

\bibitem{Champion2019-yh}
K.~Champion, B.~Lusch, J.~N. Kutz, and S.~L. Brunton, ``\BIBforeignlanguage{en}{Data-driven discovery of coordinates and governing equations},'' \emph{\BIBforeignlanguage{en}{Proc. Natl. Acad. Sci. U. S. A.}}, vol. 116, no.~45, pp. 22\,445--22\,451, Nov. 2019.

\bibitem{Chen2021-za}
J.~Chen, X.~Feng, L.~Jiang, and Q.~Zhu, ``State of charge estimation of lithium-ion battery using denoising autoencoder and gated recurrent unit recurrent neural network,'' \emph{Energy}, vol. 227, p. 120451, Jul. 2021.

\bibitem{Wei2022-zq}
M.~Wei, M.~Ye, Q.~Wang, {Xinxin-Xu}, and J.~P. Twajamahoro, ``Remaining useful life prediction of lithium-ion batteries based on stacked autoencoder and gaussian mixture regression,'' \emph{Journal of Energy Storage}, vol.~47, p. 103558, Mar. 2022.

\bibitem{Sulzer2021-ao}
V.~Sulzer, P.~Mohtat, S.~Pannala, J.~B. Siegel, and {others}, ``Accelerated battery lifetime simulations using adaptive inter-cycle extrapolation algorithm,'' \emph{Journal of The Electrochemical Society}, 2021.

\bibitem{Bocca2020-df}
A.~Bocca and D.~Baek, ``Optimal {Life-Cycle} costs of batteries for different electric cars,'' in \emph{2020 {AEIT} International Conference of Electrical and Electronic Technologies for Automotive ({AEIT} {AUTOMOTIVE})}, Nov. 2020, pp. 1--6.

\bibitem{Kevorkian1996-qm}
J.~K. Kevorkian and J.~D. Cole, \emph{\BIBforeignlanguage{en}{Multiple scale and singular perturbation methods}}, 1996th~ed., ser. Applied mathematical sciences.\hskip 1em plus 0.5em minus 0.4em\relax New York, NY: Springer, May 1996.

\bibitem{Khalil2002-zn}
H.~K. Khalil, \emph{Nonlinear systems}.\hskip 1em plus 0.5em minus 0.4em\relax Prentice-Hall, 2002.

\bibitem{Wiewel2019-ap}
S.~Wiewel, M.~Becher, and N.~Thuerey, ``\BIBforeignlanguage{en}{Latent space physics: Towards learning the temporal evolution of fluid flow},'' \emph{\BIBforeignlanguage{en}{Comput. Graph. Forum}}, vol.~38, no.~2, pp. 71--82, May 2019.

\bibitem{Gu2018-dq}
J.~Gu, Z.~Wang, J.~Kuen, L.~Ma, A.~Shahroudy, B.~Shuai, T.~Liu, X.~Wang, G.~Wang, J.~Cai, and T.~Chen, ``Recent advances in convolutional neural networks,'' \emph{Pattern Recognit.}, vol.~77, pp. 354--377, May 2018.

\bibitem{Kiranyaz2021-hy}
S.~Kiranyaz, O.~Avci, O.~Abdeljaber, T.~Ince, M.~Gabbouj, and D.~J. Inman, ``{1D} convolutional neural networks and applications: A survey,'' \emph{Mech. Syst. Signal Process.}, vol. 151, p. 107398, Apr. 2021.

\bibitem{Chung2014-yo}
J.~Chung, C.~Gulcehre, K.~Cho, and Y.~Bengio, ``Empirical evaluation of gated recurrent neural networks on sequence modeling,'' \emph{NIPS 2014 Workshop on Deep Learning}, Dec. 2014.

\bibitem{Hochreiter1997-uf}
S.~Hochreiter and J.~Schmidhuber, ``\BIBforeignlanguage{en}{Long short-term memory},'' \emph{\BIBforeignlanguage{en}{Neural Comput.}}, vol.~9, no.~8, pp. 1735--1780, Nov. 1997.

\bibitem{Pozzato2022-ad}
G.~Pozzato, A.~Allam, and S.~Onori, ``\BIBforeignlanguage{en}{Lithium-ion battery aging dataset based on electric vehicle real-driving profiles},'' \emph{\BIBforeignlanguage{en}{Data Brief}}, vol.~41, p. 107995, Apr. 2022.

\bibitem{Werbos1990-uu}
P.~J. Werbos, ``Backpropagation through time: what it does and how to do it,'' \emph{Proc. IEEE}, vol.~78, no.~10, pp. 1550--1560, Oct. 1990.

\end{thebibliography}

\end{document}